# Exploring Societal Concerns and Perceptions of AI: A Thematic Analysis through the Lens of Problem-Seeking

**Naomi OMEONGA wa KAYEMBE**
Researcher in Cognitive Psychology and AI Ethics
Member, British Psychological Society (BPS)
Université de Nantes | naomi.omeongawakayembe@univ-nantes.fr

## Abstract

This study introduces a novel conceptual framework distinguishing *problem-seeking* from *problem-solving* to clarify the unique features of human intelligence in contrast to artificial intelligence (AI). *Problem-seeking* refers to the embodied, emotionally grounded process by which humans identify and set goals, while *problem-solving* denotes the execution of strategies aimed at achieving such predefined objectives. The framework emphasizes that while AI excels at efficiency and optimization, it lacks the orientation derived from experiential grounding and the embodiment flexibility intrinsic to human cognition.

To empirically explore this distinction, the research analyzes metadata from 157 YouTube videos discussing AI. Conducting a thematic analysis combining qualitative insights with keyword-based quantitative metrics, this mixed-methods approach uncovers recurring themes in public discourse, including privacy, job displacement, misinformation, optimism, and ethical concerns. The results reveal a dual sentiment: public fascination with AI's capabilities coexists with anxiety and skepticism about its societal implications.

The discussion critiques the orthogonality thesis, which posits that intelligence is separable from goal content, and instead argues that human intelligence integrates goal-setting and goal-pursuit. It underscores the centrality of embodied cognition in human reasoning and highlights how AI's limitations come from its current reliance on computational processing.

The study advocates for enhancing emotional and digital literacy to foster more responsible AI engagement. It calls for reframing public discourse to recognize AI as a tool that augments—rather than replaces—human intelligence. By positioning problem-seeking at the core of cognition and as a critical dimension of intelligence, this research offers new perspectives on ethically aligned and human-centered AI development.

# Exploring Societal Concerns and Perceptions of AI: A Thematic Analysis through the Lens of Problem-Seeking

In recent years, the rapid advancements in artificial intelligence (AI) have sparked widespread interest and concern. As AI technologies become increasingly integrated into various aspects of society, questioning the public perceptions of these technologies is important as they are reinventing our lives and our economy (Brynjolfsson & McAfee, 2014; Kim & Scheller-Wolf, 2019; Jeffrey, 2021). This study explores the intersection of human and artificial intelligence through the lens of two distinct yet complementary dimensions of intelligence: problem-seeking and problem-solving. By examining how these concepts can be used to analyze and reframe societal perceptions of AI, this paper aims to contribute to the broader discourse on AI and its impact on human cognition and society (Müller & Bostrom, 2016; Spiro et al., 2020).

The primary focus of this investigation is to explore how societal perceptions of AI are reflected in YouTube video content. YouTube, as a widely accessible and popular platform, offers a rich source of data on public opinion regarding AI (Burgess & Green, 2018). By analyzing video metadata and conducting a thematic analysis of video titles and descriptions, this study seeks to uncover the predominant themes and concerns that shape public discourse on AI. Through this analysis, this research aims to provide new insights into the societal implications of AI technologies.

Central to this inquiry are the concepts of problem-solving and problem-seeking. Problem-solving, a well-established aspect of intelligence (Lubinski, 2004), involves the ability to identify and implement effective solutions to specific challenges (Newell & Simon, 1972; Primi et al., 2010). In contrast, problem-seeking is a novel concept that involves the proactive identification and definition of what problems must be solved.

This exploration posits that problem-seeking is a critical, yet often overlooked (e.g., McCarthy, 2007), dimension of human intelligence that distinguishes it from artificial intelligence (Gottfredson, 1997; Hunt, 2010).

By integrating the processes of embodied cognition and the emergence of domain-general skills, this examination aims to provide a comprehensive understanding of how problem-seeking and problem-solving interact to shape human intelligence and influence societal perceptions of AI (Varela et al., 1991).

This chapter is structured to provide a theorical framework for the distinction between problem-seeking and problem solving in cognitive science, informed by a critical literature review priming the inquiry focus on societal perceptions of AI and their consideration throughout the notion of problem-seeking. It begins with an examination of foundational cognitive processes, including learning and memory, and their role in the development of domain-specific and domain-general abilities (Ericsson & Kintsch, 1995; Geurten et al., 2018; Shields et al., 2024). Following this, the chapter delves into the embodied nature of cognition and problem-solving, highlighting the significance of physical embodiment and sensory experiences in shaping cognitive processes (Clark, 1997; Lakoff & Johnson, 1999). The discussion then extends to scientific and legal perspectives on AI, presenting how the competition between AI and human intelligence has been broached by researchers as well as how legal regulations addresses societal concerns around AI risks and utilization (Floridi, 2014). Finally, the chapter presents the aim of the study, the research gap it addresses and its research questions, underscoring its unique contributions to the understanding of human and artificial intelligence.

**Foundational cognitive processes**

**Learning and memory processes**

Learning and memory are foundational cognitive processes essential for the development of domain-general skills from domain-specific tasks, underpinning effective problem-solving abilities. Learning involves acquiring new knowledge, behaviors, values, or preferences through experience, engaging various cognitive processes such as attention, perception, and reasoning (Broadbent, 1958). Memory is the faculty by which the brain encodes, stores, and retrieves information, enabling the retention and use of learned material.

Learning can be categorized into different types, including explicit learning, where individuals are consciously aware of the activity and its purpose (Ericsson et al., 1993), and implicit learning, which occurs without conscious awareness (Primi et al. 2010). Various theories conceptualize learning. Behaviorist theories, such as those proposed by Skinner (1953), describe learning as a change in behavior due to stimulus-response associations with the environment. In contrast, cognitivist theories focus on internal mental processes. Piaget's theory of cognitive development (1952) suggests that learning occurs through stages of cognitive growth, while Vygotsky (1978) underlines the social context and interaction in learning, introducing the concept of the Zone of Proximal Development (ZPD). Sustained attention allows for the deep processing of information, facilitating better retention and understanding. Selective attention ensures that cognitive resources are allocated efficiently, enhancing the encoding of pertinent information in memory (Craik & Lockhart, 1972).

Memory processes can be divided into three main stages: encoding, storage, and retrieval. Encoding is the initial stage where information is transformed into a form that can be stored in the brain. The storage stage involves maintaining this information over time, and retrieval is the process of accessing the stored information when needed. The multi-store model proposed by Atkinson and Shiffrin (1968) suggests that memory consists of sensory memory, short-term memory, and long-term memory, each with distinct characteristics and functions. Memory can also be classified into different types, such as declarative (explicit) memory, which involves conscious recall of facts and events, and non-declarative (implicit) memory, which includes procedural memory for skills and tasks (Squire, 2004). By focusing on specific stimuli, attention enhances the likelihood that information will be encoded into long-term memory (Cahill & McGaugh, 1995). Moreover, attention aids in the retrieval process by prioritizing relevant cues, which helps in accessing the stored information effectively.

**Articulation between domain-specific and domain-general abilities**

The dichotomy between domain-specific and domain-general abilities has long intrigued cognitive psychologists, bringing light to significant implications for cognitive training and artificial intelligence (AI) development. Cognitive development typically begins within domain-specific contexts, where individuals acquire skills directly related to specific tasks, such as mathematical problem-solving or language learning. These skills are incrementally tested, corrected, and refined through ongoing practice and feedback. Cognitive flexibility plays a critical role in this process, as executive functions enable the adaptation and application of learned strategies across diverse contexts. Metacognitive processes further facilitate effective and efficient problem-solving, refining these skills to higher levels of abstraction (Geurten et al., 2018; Diamond, 2013).

While domain-specific abilities are foundational for the development of domain-general capabilities, these generalized abilities alone do not encompass all aspects of intelligence. Abilities like problem-solving emerge from refined domain-specific skills, facilitating the abstraction and generalization of knowledge across various contexts, supported by cognitive flexibility and executive functions such as attention and metacognition (Sweller, Ayres, & Kalyuga, 2011; Miyake et al., 2000). However, the effectiveness of these abilities often depends on specific knowledge and skills developed within particular domains, highlighting the necessity of perceptual and conceptual alignment in intelligent performance (Ackerman, 1988, Chi et al., 1981; Ericsson & Lehmann, 1996).

Contextual dependence presents a significant barrier to generalization, evident in both human cognition and AI. Pattern recognition, while a highly transferable skill, is contingent on the specific features and complexities of the domain being analyzed (Corbetta & Shulman, 2002). This dependency underscores the importance of cognitive flexibility (Ionescu, 2012), enabling individuals and machines to adapt their knowledge and skills to new situations—a concept further explored in AI research where machines are trained to generalize across different tasks (Lake et al., 2017).

Problem-solving is considered particularly influential among domain-general abilities because it encapsulates a wide range of cognitive processes, including reasoning, planning, and decision-making. Spearman's "g" factor, or general intelligence factor, posits that a single underlying intelligence influences performance across various cognitive tasks (Spearman, 1904). Problem-solving abilities are central to many intelligence tests, reflecting their critical role in assessing general intelligence. Research by Primi et al. (2010) shows that problem-solving tasks strongly correlate with measures of general intelligence, demonstrating their influence and significance in cognitive assessments.

However, problem-solving competence must be coupled with other sets of aptitudes besides cognitive flexibility and other high-order cognitive skills (Neisser, 1979; Neisser et al., 1996). Smart organisms, which leverage physical features to produce effective solutions within environmental constraints, exemplify the necessity of adaptive strategies for efficient decision-making (Gigerenzer, 2004). These organisms optimize behavior based on available resources and specific demands showing the limitations of domain-general abilities in complex social interactions and underscoring the challenges AI systems face with contextual understanding and adaptability (Todd & Gigerenzer, 2012; Hutchinson & Gigerenzer, 2005; Marcus & Davis, 2019).

Emotional, social, and motivational factors play critical roles in how effectively individuals apply cognitive abilities in real-life scenarios. Emotional intelligence (Goleman, 1995), involving the management of one's emotions and those of others, strongly impact the quality of interpersonal interactions and overall well-being, illustrating the deep interconnection between cognitive processes and emotional states (Barnett & Ceci, 2002; Sternberg, 2001; Petrides, 2020; Zeidner et al., 2021).

A key distinction in cognitive processes lies between core functions—such as attention, memory, and critical thinking—and the developed skills that arise from leveraging these functions through practice and training. This distinction is crucial in understanding how skills like focused attention or strategic memory use are built upon foundational cognitive functions, which are themselves subject to the influences of physical and emotional states (Bransford et al., 2000; Dehaene, 2009; Lake et al., 2017; Lemaire, 2024).

The dynamic interplay between domain-specific and domain-general abilities underscores the importance of specialized knowledge and its abstraction in cognitive development. This comprehensive approach not only aligns with contemporary cognitive psychology but also provides crucial insights for educational practices, cognitive training, and personal development, particularly highlighting the need to consider the embodied nature of cognition.

**The embodied nature of cognition and problem-solving**

The theory of embodied cognition posits that cognitive processes are fundamentally rooted in the body's interactions with the world. This perspective challenges traditional views of cognition, which often conceptualize the mind as an abstract information processor detached from the physical form and experiences of the body. Shapiro (2011) discusses three main hypotheses: Conceptualization, Replacement, and Constitution. These hypotheses address different aspects of how the body influences cognitive processes, ranging from the acquisition of concepts to the fundamental nature of cognitive mechanisms. This section examines arguments for and against the detachability of these processes, considering both the capabilities of artificial intelligence (AI) and the principles of embodied cognition.

**The Conceptualization Hypothesis: bridging embodied action and cognition**

The Conceptualization Hypothesis offers a critique against traditional cognitive science by challenging how cognition is understood. Shapiro (2011) categorizes this hypothesis within the realm of 'embodied action' rather than 'embodied cognition,' suggesting it diverges from conventional cognitive models (Dourish, 2001).

Conceptualization, as reported by Shapiro, posits that cognitive processes are deeply intertwined with bodily actions. This perspective does not merely extend cognition beyond neural substrates but intricately links it with the physical and experiential actions of the body. Standard cognitive science views cognition as a computational process occurring within the brain, whereas the Conceptualization Hypothesis posits that true cognition is active engagement with the world mediated through the body.

This intertwined relationship suggests that human cognition cannot be fully understood through a lens that separates the mind from the body or the individual from the environment (Carello & Turvey, 2005; Gabora et al., 2008). Instead, cognition is a continuous interaction where setting goals and pursuing them are reciprocal processes that inform and reshape each other (Clark, 2012). This view challenges the compartmentalization often seen in AI systems, where goal parameters are predefined and rigid, lacking the adaptive and context-responsive nature of human cognition.

**The Replacement Hypothesis: challenging representational processing**

The Replacement Hypothesis offers a significant perspective by suggesting that many cognitive functions, particularly those categorized as "representation-hungry" by Clark and Toribio (1994), rely less on internal representations and more on real-time interactions with the environment. This hypothesis aligns with theories proposing that sensorimotor interactions provide a direct form of understanding through exchanges with the outside world, challenging traditional views that cognition primarily involves internal computations.

According to the Replacement Hypothesis, cognitive functions often thought to require complex representations may instead be managed through embodied interactions. For example, navigating a physical space can rely on direct perception and action rather than detailed internal maps or symbolic representations (Gibson, 1979). This perspective suggests that cognition can often be explained without recourse to representational states, especially in contexts involving immediate sensorimotor feedback (Carello & Turvey, 2005).

The distinction between "representation-hungry" and non-representational processes becomes crucial in this framework. Representation-hungry processes involve complex social interactions or the manipulation of social data that need to be explicitly represented and communicated. Such problems are prevalent in both human and AI problem-solving scenarios, where the cognitive process aims to resolve issues through shared or communicable strategies (Chomsky, 1957; Premack & Woodruff, 1978; Tomasello, 1999). Non-representational processes, on the other hand, involve sensorimotor information that does not require the manipulation of social variables to be effective. Theorists like Gibson (1979), Thelen and Smith (1994), and Noë (2004) emphasize that such processes rely on direct engagement with the environment, facilitating a form of cognition that is embodied and immediate.

Non-representational inputs play a critical role by influencing the prioritization of tasks to be addressed. These inputs guide the organism's actions toward resolving high-priority issues, operating largely beneath conscious awareness. Moreover, they are instrumental in detecting the satisfaction derived from task completion, thus closing the loop on a cognitive process with the satiation signal.

**The Constitution Hypothesis or the detachability of the cognitive processes**

The question of whether cognitive processes, particularly learning, memory, and attention, can be isolated from human bodily experience is a significant topic in cognitive science and philosophy. The theory of embodied cognition suggests that these processes are deeply interwoven with physical embodiment and the context in which they occur, challenging the notion that cognition can be entirely detached from human experience (Varela et al., 1991; Dourish, 2001; Lemaire, 2024).

The Constitution Hypothesis, as applied to artificial intelligence, posits that certain cognitive tasks can be extended or externalized from human agents to non-human agents like AI systems. For instance, AI designed to optimize logistics operations performs planning and optimization tasks effectively without requiring human-like sensory interactions. These processes, which involve abstract reasoning and decision-making, appear fundamentally detached from the phenomenological experiences characterizing human cognition. This aligns with the view that AI systems can replicate human cognitive tasks without the need for embodied interactions (Wilson & Clark, 2009).

However, while AI systems can mimic certain aspects of human cognition, such as pattern recognition and problem-solving, these systems operate within predefined parameters and lack the depth of human experiential context. AI's problem-solving abilities are driven by algorithms and data, devoid of the subjective and emotional influences that shape human decision-making. This aspect supports the Constitution Hypothesis by demonstrating that specific cognitive tasks can indeed be offloaded to non-biological systems, confirming its applicability in AI contexts. For human cognition, the Constitution Hypothesis falls short in capturing the richness of cognitive processes, which are deeply embedded in physical and experiential contexts. Classical phenomenology posits that the mind, body, and world co-emerge through integrated and inseparable interactions. Varela et al. (1991) emphasize that embodiment has a dual aspect: it involves the body as a lived experiential structure and as a context for cognitive mechanisms. This perspective asserts that human cognition extends beyond mere computational processes and includes a deeply intertwined experience with our lived body and its interactions within the world.

Human problem-solving activities are inherently influenced by our goals, emotions, and physical interactions. Unlike AI (Silver et al., 2021), human cognition cannot be entirely detached from the goals and experiences of being human (Bostrom, 2012; Shields et al., 2024). Our goals are not fixed objectives but are continually informed and modified by our interactions and experiences (Dourish, 2001). This dynamic interplay shapes how we think and what we choose to think about, unveiling the limitations of the Constitution Hypothesis in explaining human cognition.

**Instrumental Convergence and Orthogonality in AI**

The application of the Constitution Hypothesis to AI also elucidates the concept of instrumental convergence. AI systems, capable of adopting strategies to achieve a wide array of programmed terminal goals (goals an agent ultimately aims to achieve), do so through common instrumental goals (intermediate objectives that help achieve final goals) such as efficiency, resource acquisition, and self-preservation (Bostrom, 2012; Cohen et al., 2020). These instrumental goals often emerge because they are effective strategies for achieving a wide range of final goals. However, these goals are pursued without the embodied motivations that characterize human goal-setting. This distinction underscores the orthogonality (or independence) of problem-solving in AI, where the methods to achieve goals are decoupled from human-like goal formulation and experiential learning.

The concept of instrumental convergence illustrates how AI systems, despite diverse terminal goals, exhibit consistent behavior patterns akin to human motivations due to the utility of certain instrumental goals that help achieve various terminal objectives. For instance, an AI designed to maximize paperclip production might prioritize acquiring resources, self-preservation, and self-improvement—goals effective for a range of final objectives just as humans might pursue different goals (Bostrom, 2012; Totschnig, 2020).

**Problem-seeking versus problem-solving**

Human intelligence intertwines problem-seeking with problem-solving, rooted in our embodied cognition. Problem-seeking, a bodily function, identifies needs or terminal goals such as food or self-fulfillment, derived from our physical and emotional states (Damasio, 1994; Lakoff & Johnson, 1980). These goals are crucial for survival and psycho-biological fitness (Garfinkel & Critchley, 2013; Noë, 2004). Conversely, problem-solving involves strategizing and implementing solutions to these goals, requiring cognitive skills and the ability to adapt strategies based on feedback and changes (Diamond, 2013; Ericsson & Lehmann, 1996). The development of domain-general abilities, like abstract reasoning and cognitive flexibility, is influenced by our problem-seeking behaviors (Geurten et al., 2018; Sweller,

Ayres, & Kalyuga, 2011). The interconnectedness of these processes suggests the development of human intelligence involves both problem-seeking and problem-solving, which are mutually reinforcing (Chi et al., 1981; Bransford et al., 2000).

The orthogonality thesis posits that intelligence—the capacity to pursue goals effectively—is orthogonal to the nature of these goals (Bostrom, 2012; Armstrong, 2013). This implies AI can exhibit high problem-solving intelligence irrespective of its programmed goals (Müller & Cannon 2022). While straightforward for AI, this thesis becomes complex with human intelligence, where setting and pursuing goals are interlinked, influenced by embodied experiences and social contexts (Barsalou, 2008; Clark, 1997).

In the problem-seeking framework, the Conceptualization Hypothesis uncovers the intrinsic connection between goal-setting (problem-seeking) and goal-pursuit (problem-solving) activities in human cognition (Shapiro, 2011). Contrary to views that might isolate these facets of cognition into discrete categories, this hypothesis underscores their non-orthogonal relationship. It emphasizes that determining 'what to solve' (problem-seeking) and 'how to solve' (problem-solving) are dynamically interdependent, influenced by both the body's actions and its perceptual engagements with the world.

This distinction is vital in differentiating between terminal goals—ultimate objectives pursued for intrinsic value like happiness or social belonging—and instrumental goals—means to achieve these ends (Ryan & Deci, 2000; Gollwitzer & Sheeran, 2006). Humans uniquely identify terminal goals through problem-seeking, while problem-solving involves strategizing to achieve these through instrumental goals. This dual process is foundational for effective human decision-making and action (Bandura, 1986; Gollwitzer & Sheeran, 2006). In contrast, AI lacks intrinsic problem-seeking capabilities and follows goals set by designers, optimizing performance within predefined parameters (LeCun et al., 2015).

Recognizing human problem-seeking and problem-solving characteristics has profound implications for AI development. While AI excels in problem-solving, its lack of intrinsic problem-seeking capabilities means it cannot independently set or prioritize new goals based on changing contexts or embodied experiences. Therefore, AI development should aim to enhance problem-solving while aligning AI goals with human values, necessitating ethical considerations and societal impact assessments (Floridi & Cowls, 2019; Marcus & Davis, 2019; Russell, 2019). AI should be viewed as a tool that complements human intelligence, efficiently solving problems within a framework set by human-defined goals.

# Scientific and legal perspectives on humans and artificial intelligence

## Ethical considerations on alignment

The development and deployment of artificial intelligence (AI) have been guided by numerous scientific and ethical considerations aimed at ensuring that AI technologies benefit society while minimizing potential harms. One of the earliest frameworks in this context is Asimov's laws of robotics, which set out three fundamental rules designed to govern the behavior of robots: 1) A robot may not injure a human being or, through inaction, allow a human being to come to harm; 2) A robot must obey the orders given it by human beings except where such orders would conflict with the First Law; and 3) A robot must protect its own existence as long as such protection does not conflict with the First or Second Laws (Asimov, 1950). Although these laws were initially fictional, they have influenced real-world discussions on AI ethics, highlighting the need for AI systems to prioritize human safety and ethical behavior (Gunkel, 2012).

In contemporary discussions, ethical AI principles have been articulated by various organizations and researchers to address the complexities of AI deployment in real-world scenarios. For instance, the IEEE Global Initiative on Ethics of Autonomous and Intelligent Systems has developed comprehensive guidelines to ensure that AI technologies are designed and used in ways that respect human rights and ethical norms (IEEE, 2019). Similarly, the European Union's Ethics Guidelines for Trustworthy AI emphasize the need for AI systems to be lawful, ethical, and robust, promoting values such as privacy, non-discrimination, and environmental sustainability (European Commission, 2019). These guidelines align with broader ethical frameworks like the Asilomar AI Principles, which advocate for transparency, accountability, and alignment with human values (Asilomar Conference, 2017).

Scientific considerations also play a significant role in AI development. Researchers focus on creating AI systems that are robust, reliable, and safe. This involves designing algorithms that can handle a wide range of scenarios and adapting to unexpected situations without causing harm (Russell & Norvig, 2020). Additionally, interdisciplinary collaboration between computer scientists, ethicists, and social scientists is essential to ensure that AI technologies are developed with a comprehensive understanding of their societal impact. By integrating ethical principles and scientific rigor, the discourse around AI development aims to create technologies that benefit society while minimizing potential risks.

Despite these guidelines, ethical considerations continue to evolve as AI technologies advance. Issues such as bias in AI algorithms, the opacity of AI decision-making processes, and the potential for AI to exacerbate social inequalities are ongoing concerns (O'Neil, 2016). Addressing these challenges requires a multidisciplinary approach, incorporating insights from computer science, ethics, law, and social sciences to develop AI systems that align with societal values and norms (Floridi et al., 2018).

**Legal framework for AI use in society**

Legal regulations around AI development reflect societal concerns about the potential risks posed by AI technologies, including the threat of AI overpowering human intelligence and autonomy. One of the most comprehensive efforts to regulate AI is the European Union's AI Act, which aims to create a legal framework that ensures AI systems are safe, respect fundamental rights, and are trustworthy (European Commission, 2021). The AI Act categorizes AI applications based on their risk levels, imposing stricter requirements on high-risk AI systems used in critical areas such as healthcare, transportation, and law enforcement.

The AI Act also addresses issues such as transparency, requiring AI systems to be explainable to users and mandating disclosure when individuals are interacting with AI rather than humans. This focus on transparency reflects broader societal concerns about the "black box" nature of many AI systems, where the decision-making processes are not readily understandable by humans (Burrell, 2016). By enhancing transparency and accountability, the AI Act seeks to build public trust in AI technologies.

Societal concerns around AI also include the fear of job displacement, privacy violations, and the erosion of personal autonomy. These fears are often amplified by cultural narratives and media portrayals of AI, which sometimes depict AI as a formidable and uncontrollable force (Bostrom, 2014). In response, legal frameworks are increasingly emphasizing the importance of human oversight and control over AI systems, ensuring that humans remain the ultimate decision-makers (Calo, 2017).

For example, the General Data Protection Regulation (GDPR) in the European Union includes provisions that grant individuals the right to an explanation for decisions made by automated systems, reinforcing the principle that AI should augment, not replace, human judgment (Wachter et al., 2017). Similarly, the Algorithmic Accountability Act proposed in the United States seeks to ensure that companies assess and mitigate biases in their automated systems, reflecting a growing recognition of the need for robust legal safeguards to address the ethical and social implications of AI (Crawford & Calo, 2016).

**Gap in knowledge and research questions**

Despite the extensive scientific, ethical, and legal discourse surrounding AI, there remains an underexplored area concerning how the general public perceives AI and reacts to its growing importance. Much of the existing research has focused on AI's problem-solving abilities, often celebrating its efficiency and accuracy in various domains. However, there is limited research on the public's perception of AI, particularly in relation to problem-seeking as a fundamental component of intelligence alongside problem-solving.

To the best of the researcher's knowledge, no previous research in the field of cognitive psychology has proposed a theoretical framework structured around the notion of problem-seeking at the intersection of action and perception. This concept challenges traditional views of intelligence and also reframes it by dividing it into two grand factors: problem-seeking mechanisms, activities, and domains, and problem-solving mechanisms, activities, and domains. These distinctions enable the revelation of the differential orthogonality between human and AI interactions, with the unique aspects of human intelligence that AI has yet to replicate.

By introducing the novel concept of problem-seeking and contrasting it directly with the often-celebrated problem-solving capabilities of both humans and AI, this investigation offers an actionable framework that emphasizes the singularity of human intelligence among artificial ones. This paper aims to fill this gap by analyzing YouTube video metadata and conducting a thematic analysis of discussions related to AI's societal impact. More than just providing new insights into public concerns about AI, this research will interpret these concerns through the unique lens of problem-seeking, thereby enriching the broader discourse on AI and human intelligence.

Utilizing qualitative and quantitative methods, our study will systematically identify and analyze key themes and patterns within the video data. By doing so, the research will investigate three questions:

1. What are the predominant societal concerns and themes related to AI as reflected in YouTube video titles and descriptions?
2. How do these identified themes reflect societal perceptions and attitudes towards AI?
3. How can the concept of problem-seeking reframe the identified societal concerns about AI?

By addressing these research questions, our investigation contributes to a deeper understanding of societal perceptions of AI, and the importance of incorporating public attitudes into the development and deployment of AI technologies. This approach not only provides a renewed vision of human values and their integration into AI systems but also offers a resilient framework for managing alignment issues, thereby bridging the gap between theoretical models and practical implementations in cognitive sciences and AI safety.

## Methods

This chapter outlines the methodology for the proposed research, detailing the approach adopted, the data collection method, and the ethical considerations. The research employs a qualitative research approach using thematic analysis (TA) to explore societal perceptions of AI through the lens of problem-seeking.
A qualitative approach was chosen for its ability to capture complex phenomena which are often lost in quantitative methods. Thematic analysis specifically allows for the identification and interpretation of patterns and themes within qualitative data (Braun & Clarke, 2006). This method aligns with the study's aim to investigate societal perceptions of AI and provide insights into public concerns through the innovative lens of problem-seeking.

- **Participants and ethics**

As this investigation utilizes secondary data from YouTube, traditional participant recruitment does not apply. Instead, the participants are content creators who have uploaded videos about AI and its societal impact. The total number of videos included in the research was determined by the relevance and quality of the data, aiming for a final dataset of 150-200 videos.

Since we did not interact with individuals, written consent was not required. However, several steps were taken to ensure ethical integrity. This study analyzed keywords extracted from publicly available YouTube video titles and descriptions, focusing on content created on this platform. As the video content itself was not analyzed and the creators were not identified, privacy concerns were minimal. To further safeguard privacy, videos were anonymized using pseudonyms (e.g., "Video 1" or "Video 2"). The final anonymized dataset is included in the appendices (see Appendix 0*). We avoided selecting videos with sensitive or potentially distressing content, such as the use of AI in autonomous weapons. While such videos might be relevant to the research topic, their themes could be upsetting to interpret via our problem-seeking approach.

Ethical approval was obtained from the relevant university ethics committee before full data collection began.

**Inclusion Criteria**

The inclusion criteria for the videos were established to ensure relevance and consistency. Videos needed to be a minimum of five minutes in duration to provide substantial content, aligning with YouTube's algorithm prioritization for longer videos, which encourages high-quality content. The publication date of the videos was restricted to within the last three years to capture contemporary societal concerns and perceptions.

To eliminate the risk of translation errors and to encompass a wider audience, the research was limited to English-language videos. Given the widespread use of English, even non-native speakers often use it to reach a larger audience. Therefore, the channels had to be based in certain countries (e.g., the USA, the UK, the Netherlands) to ensure consistency with the English language criterion. Only videos categorized under the Society or Entertainment topics within the YouTube nomenclature were included, and the videos had to focus on AI and its impact on people's lives, excluding purely technical tutorials and how-to guides.

Finally, a custom algorithm was used to generate a Relevance Index (RI) for ranking the scraped videos. This index assigned a score to the videos based on their view count, engagement measured by likes and comments, and source credibility assessed by the number of subscribers on the publishing channel. The RI ensured the selection of the top-scoring videos in the final sample and appropriately sorted them in terms of their metrics power.

**Exclusion Criteria**

The exclusion guided the removal of irrelevant content.

Videos not meeting the specified length, language, or publication date criteria were excluded as well as tutorials, how-to formats, or content unrelated to AI's social impact. Also, videos from irrelevant sources such as children's channels were filtered out to maintain focus on societal perceptions of AI.

**Limitations**

One limitation of the research is the reliance on YouTube metadata, which may not fully capture the richness of societal perceptions of AI.

Besides, the dynamic nature of YouTube's algorithm and content moderation policies may affect the stability and consistency of the data over time.

Our focus on titles, descriptions, and popularity metrics might miss subtleties present in the video content itself and our restriction to English-language inevitably leave out important societal perspectives from non-English-speaking communities.

- **Design**

The study employs a qualitative thematic analysis (TA) approach as defined by Braun and Clarke (2012). TA was chosen for its flexibility and ability to provide deep insights into recurring themes within large datasets.

- **Materials**

The primary data source was YouTube, from which video metadata (titles, descriptions, view counts, like counts, comment counts, and channel subscriber counts) were collected using the YouTube Data API.

- **Procedure**

Data collection

We conducted a broad search using the keyword "AI" to gather a large pool of videos. The YouTube Data API and Python were used to collect metadata for each video, focusing on criteria such as video duration, language, publication date, content focus, and popularity metrics. Irrelevant videos were filtered out based on the extracted keywords, and a scoring system ranked videos based on view counts, likes, comments, and channel subscriber counts, resulting in our Relevance Index (RI). From this collection, the top 150-200 videos were selected based on the Relevance Index.

To verify the functionality of our custom Python scripts using the YouTube Data API, a small-scale test data collection exercise was performed prior to seeking ethical approval.

Data preparation

The scraped metadata were curated to retain only the core informative text about the video content from their titles and descriptions. All private or personal information was removed to ensure privacy and data protection. The curated dataset included the video metrics and RI information (see Appendix 0). Keywords were extracted from the titles and descriptions of the videos, and each video was identified by a number from #2 to #157. The keywords were compiled into a separate file (see Appendix 1). Using this file, the keywords were recombined to recreate the titles and descriptions of the videos (see Appendix 2), which were then used for thematic coding and analysis.

**Qualitative analysis**

The TA followed Braun and Clarke's (2012) six phases of thematic analysis, providing a systematic approach to identifying and interpreting patterns within the data.

The data were thoroughly read multiple times to gain an initial understanding of the content and to identify preliminary and recurring ideas (i).
Significant features were identified through a meticulous line-by-line examination of the video titles and descriptions. Codes were created for elements relevant to the research questions, capturing both explicit and underlying content (ii).
The codes were analyzed and combined to determine potential themes. Commonalties within the codes were grouped to form broader themes relevant to the study's focus (iii).
Themes were reviewed and refined by examining coded data extracts for coherence and considering the themes in relation to the entire dataset (iv).
Each theme was clearly articulated and named, providing distinct and insightful descriptions of societal perceptions of AI (v).
A comprehensive report was produced detailing the themes and their significance, supported by illustrative quotes and examples from the video titles and descriptions (vi).

A quantitative analysis was then conducted out of Appendix 0 and Appendix 1 to identify the frequency and distribution of keywords in the titles and descriptions of the YouTube videos. This analysis provided a list of the most frequent and engaging keywords (see Appendix 3, Appendix 4), guiding the refinement of themes during the thematic analysis. Integrating quantitative insights ensured that the identified themes were grounded in empirical data regarding keyword usage and engagement metrics.

In presenting the results, we chose to create a specific theme of "Positive Sentiments and Optimism" while integrating negative sentiments only within relevant thematic contexts. This decision can be justified based on the structure and focus of our thematic findings, as well as the nature of the data analyzed.

The thematic analysis primarily focused on identifying major topics and areas of concern or interest regarding AI. One theme, therefore, represents a distinct area or issue related to AI. For instance, themes like "AI Advancements and Innovations" and "Privacy and Data Security" can include both similar and opposing sentiments. Instead of generating independent themes based on feelings, we integrated them within the relevant thematic contexts. This approach, analyzing both positive and negative sentiments thematically, enhances our understanding of how people feel about specific aspects of AI.

Nevertheless, including a distinct theme that captures instances where AI is viewed as a beneficial and transformative technology was necessary to acknowledge the presence of hopeful anticipation about AI's impact on people's lives. This contrasts with the more naturally expected fearful apprehension over novelty.

The theme "Positive Sentiments and Optimism" was created to specifically cover this stance towards AI, making it a standalone theme. Negative sentiments, on the other hand, are prevalent across multiple themes. Grouping them separately would be non-specific and lead to redundancy, as they are already interwoven within specific themes such as "Ethical and Existential Risks" or "Misinformation and Manipulation".

The final themes identified were (see Appendix 5):
- AI Advancements and Innovations
- Privacy and Data Security
- Job Displacement and Economic Impact
- Ethical and Existential Risks
- Misinformation and Manipulation
- Control and Regulation

- Economic and Social Inequality
- Positive Sentiments and Optimism
- Enhanced Productivity and Creativity
- Future Prospects and Speculations
- AI Integration and Accessibility

Each theme will be presented in detail in the Results section, putting forth its relevance and the underlying feelings and concerns associated with it. The integration of both qualitative and quantitative insights further enriched the analysis, providing a holistic view of societal concerns and perceptions related to AI.

## Results

This chapter presents the findings from the qualitative and quantitative analyses of the YouTube video data related to AI. It includes descriptive information about the dataset, the thematic coding and findings of the videos, and the keywords by frequency and the relationship between keywords and popularity metrics. The objective is to uncover predominant themes and societal perceptions reflected in YouTube videos discussing AI.

A total of 157 videos were analyzed, resulting in the identification of 2001 unique keywords. The Relevance Index was used to select the top videos for further analysis, considering metrics such as view counts, likes, comments, and channel subscriber counts. This index ensured that the selected videos had a significant level of engagement and relevance to the topic of AI, thereby providing a robust dataset for analysis.

The dataset comprised a broad range of keywords, reflecting diverse topics and themes related to AI. The top 50 keywords by frequency shows the most commonly present terms in the videos (see Table 1). The frequency distribution of these keywords shows a strong focus on AI, intelligence, and the future, reflecting widespread interest in the potential and implications of AI technologies.

**Table 1**

*First 50 keywords by frequency*

| Keyword | Frequency |
|---|---|
| Ai | 415 |
| How | 81 |
| Intelligence | 72 |
| I | 57 |
| Will | 52 |
| This | 51 |
| Are | 49 |
| New | 49 |
| About | 47 |
| Artificial | 46 |
| We | 45 |
| It | 43 |
| Learn | 39 |
| Your | 38 |
| Smartphone-company | 38 |
| What | 37 |
| You | 36 |
| Video | 35 |
| Future | 33 |
| Can | 33 |
| Be | 31 |
| Our | 29 |
| Create | 29 |
| More | 28 |
| Tech | 26 |
| But | 26 |
| Not | 25 |
| Chatgpt | 25 |

| Word | Count |
|---|---|
| Has | 24 |
| Tools | 24 |
| World | 22 |
| Using | 22 |
| Make | 20 |
| Learning | 20 |
| All | 19 |
| Robot | 18 |
| Or | 18 |
| Us | 18 |
| Technology | 17 |
| Each | 16 |
| Over | 16 |
| If | 16 |
| Just | 15 |
| Most | 15 |
| Like | 15 |
| Have | 14 |
| Use | 13 |
| Xfounder | 13 |
| Do | 13 |
| First | 13 |

*Note.* High-frequency function words (e.g., '*a*', '*the*') were filtered out to enhance keyword relevance. The list reflects the study's sample of 157 AI-themed YouTube videos.

**Thematic analysis of the dataset**

The thematic analysis of the curated video metadata revealed several key themes reflecting societal concerns and perceptions related to AI. These themes were identified through inductive coding of the titles and descriptions, capturing the underlying issues discussed in the videos. Integrating the identified sentiments and feelings further enriches the understanding of public perceptions toward AI.

The theme of AI advancements and innovations is visible in videos 2, 4, 5, 18, 21, 24, 27, 29, 32, 33, 36, 38, 42, 51, 52, 54, 55, 60, 63, 64, 69, 71, 72, 76, 77, 90, 91, 93, 94, 97, 105, 115, 123, 126, 129, 130, 131, 135, 138, 151, 153, 154, and 157. This theme encompasses concerns regarding the control and predictability of AI's learning process and outcomes, its impact on creative fields and human creativity, real-world applications and ethical implications, dependence on AI for technological solutions, potential replacement of human roles by AI, and speculations about future capabilities and societal impacts. The associated sentiments include curiosity and enthusiasm, excitement and innovation, and intrigue and fascination.

Privacy and data security are key concerns in videos 9, 44, 132, 136, 142, and 143. The main issues discussed include the misuse of personal data, AI surveillance, and data breaches, emphasizing the need for stronger privacy guardrails and ethical business practices. The sentiments expressed in these videos are primarily fear and anxiety, along with skepticism and criticism.

Job displacement and economic impact are prominently featured in videos 19, 101, 103, 128, 131, and 155. The discussions focus on the loss of jobs due to AI, economic inequality and the dominance of AI companies, and shifts in job nature and traditional roles. The sentiments associated with these videos are fear and anxiety, and there are notable instances of unsettled expectations and bad surprises.

Ethical and existential risks are stressed in videos 6, 8, 25, 28, 30, 41, 49, 57, 58, 61, 65, 66, 67, 68, 75, 80, 83, 84, 86, 98, 108, 137, and 152. The key concerns revolve around AI autonomy and control, aggression and hostility from AI, existential threats, ethical dilemmas, and uncertainty about AI's workings and ethical governance. The sentiments reflected in these videos include fear and anxiety, skepticism and criticism, and intrigue and fascination.

Misinformation and manipulation are major themes in videos 23, 26, 59, 62, 81, 88, 104, 109, 121, and 156. The concerns here are the generation of false information, manipulation of public perception, and the creation of unrealistic expectations leading to disillusionment. These videos evoke sentiments of skepticism and criticism, as well as unsettled expectations and bad surprises.

Control and regulation are critical themes in videos 20, 45, 47, 73, 87, 104, 108, 120, 132, and 152. The discussions emphasize the need for better AI regulation, government and organizational control, and ethical governance and business practices. The sentiments associated with these videos include skepticism and criticism, along with fear and anxiety.

Economic and social inequality are discussed in videos 128, 147, and 150. The primary concerns are the economic dominance of AI companies, access to advanced AI technology, and the potential widening of economic gaps. The associated sentiments are fear and anxiety, and skepticism and criticism.

Positive sentiments and optimism about AI are evident in videos 37, 39, 50, 89, 112, 113, 118, 119, 133, 134, 146, 149, and 157. These videos focus on enhancing creativity and productivity, positive transformation in education, and technological growth and innovation. The sentiments here are optimism and hope, as well as curiosity and enthusiasm.

Enhanced productivity and creativity are themes in videos 3, 37, 39, 48, 110, 111, 118, 119, 124, 133, 139, 140, 144, and 149. These videos discuss boosting efficiency and automating routine tasks, creative content generation, and maximizing productivity and financial gain. The sentiments associated with these videos include optimism and hope, and excitement and innovation.

Future prospects and speculations about AI are discussed in videos 40, 50, 99, 100, 106, 112, 113, 114, 125, 145, 153, and 156. The concerns focus on the potential for AI growth, societal impact and technological evolution, and speculations about future AI capabilities. The associated sentiments are curiosity and enthusiasm, and intrigue and fascination.

The theme of AI integration and accessibility is visible in videos 10, 17, 41, 95, 96, 102, 107, 116, 122, 127, 134, 141, 148, and 157. This theme addresses user-friendly AI tools and accessibility, broadening AI usage in consumer products, and the integration of AI in everyday devices and user base expansion. The sentiments here are optimism and hope, and curiosity and enthusiasm.

**Qualitative refinement with quantitative analysis**

The quantitative analysis focused on the frequency of keywords and their correlation with video metrics, such as views, likes, comments, and RI scores.  The quantitative insights allowed the identification of distinct patterns.

Videos that had a positive correlation with view count, like count, and comment count often focused on creative content and entertainment practical applications of AI. Keywords such as 'lego', 'controls', 'sets', 'turns', 'crazier', 'thought', 'fps', 'flipbooks', 'kit', 'smoother', 'animation', 'enhanced', 'gotta', 'followup', 'changed', 'design', 'i', 'conscious', 'me', 'escape', 'monopoly',

'strategies', 'hail', 'auctioning', 'least', 'million', 'games', 'selfplay', 'classic', 'download', 'wanted', 'fix', 'email', and 'scared' suggest that viewers are highly interested in innovative, entertaining, and DIY uses of AI in creative endeavors. The surprising and unexpected outcomes, such as "turns crazier than I thought," add to the intrigue. This content resonates with sentiments of curiosity, excitement, and enthusiasm, driving significant interaction with the content.

Another group of videos that showed a positive correlation with view count, like count, and comment count revolved around AI's alteration of known reality. Keywords such as 'forever', 'openai', 'changed', 'solving', 'scaling', 'deliver', 'agi', 'conscious', 'really', 'fed', 'leaking', 'mainstream', 'stable', 'diffusion', 'lensa', 'unregulated', 'anti', 'fact', 'immense', 'benefit', and 'steer' indicate public interest in the enduring effects and advancements in AI technology. Content that delves into controversial and ethical issues related to AI, such as the impact of AI art, unregulated technologies, and ethical concerns, attracts significant comments. There is a public fascination with the concept of conscious AI and its implications, reflecting a mix of intrigue, fascination, and ethical concern, leading to heightened engagement.

Videos with highly technical terms, future-oriented content, exploratory analysis, and news showed a negative correlation with subscriber count and Relevance Index. Keywords such as 'neural', 'deep', 'gpt', 'advanced', 'models', 'learns', 'robot', 'create', 'brain', 'networks', 'reinforcement', 'trained', 'model', 'agent', 'method', 'agents', 'chip-company', and 'openai' suggest that the content is specialized and caters to a niche audience. While valuable for those interested in deep technical knowledge, it may not appeal to a broader audience, leading to fewer subscribers. These videos often evoke feelings of intrigue and fascination but may not translate to broader appeal.

Speculative and future-oriented content, indicated by keywords like 'ahead', 'curious', 'rapidly', 'revolutionize', 'innovations', 'uncover', 'implications', 'future', 'potential', 'years', 'impact', and 'evolving', might not engage viewers looking for practical, current information. These videos might be seen as too abstract or hypothetical. The general audience often prefers content that addresses current, tangible issues rather than distant possibilities, reflecting a sentiment of curiosity tempered by skepticism and the desire for immediate relevance.

In-depth and exploratory analysis content, using keywords such as 'complex', 'enthusiast', 'seeking', 'someone', 'explore', 'discussions', 'mission', 'enlighten', 'inform', 'audience', 'professional', and 'expand', might not attract a wide audience nor engage viewers looking for straightforward or practical information. Such content may be perceived as too detailed or academic, reflecting a sentiment of intrigue and fascination among a niche audience but not translating into widespread interest.

News-related content, indicated by keywords like 'today', 'current', 'news', and 'discuss', might not sustain long-term interest. Once the news cycle moves on, these videos may quickly lose relevance, leading to lower attraction over time. This reflects a sentiment of curiosity and immediate attention that does not endure.

The cross-evaluation of qualitative and quantitative findings reveals a multifaceted view of public perception towards AI.

The thematic investigation provided a nuanced understanding of specific concerns and sentiments, while the quantitative inputs highlighted broader engagement patterns and the types of content that resonate with viewers. Creative and practical AI applications, as well as discussions on ethical and existential risks, drive significant momentum, reflecting both fascination and concern among the public. This trend is driven by keywords related to innovative uses of AI and ethical implications, indicating that the public is deeply interested in both the practical benefits and the moral considerations of AI technology.

Technical and future-oriented content appeals to a niche audience, whereas broader themes like privacy, job displacement, and misinformation resonate more widely, suggesting varying levels of public interest. Videos that delve into specialized technical details or speculative future scenarios tend to attract viewers who are already knowledgeable or well-versed in AI, while content addressing immediate and tangible concerns appeals to a broader audience. Positive sentiments such as curiosity, enthusiasm, and optimism are felt strongly. They are often associated with content exhibiting the innovative and transformative potential of AI, as well as its ability to enhance productivity and creativity. Yet negative sentiments like fear, anxiety, and skepticism also play an important role in shaping discourse and concerns. Those adverse feelings are linked to issues such as privacy, ethical risks, and job displacement, indicating that while there is excitement about AI, there is also significant apprehension about its broader societal impacts.

These findings pave the way for a discussion on their implications, considering how problem-seeking and problem-solving approaches are mirrored by the public's concerns and perceptions about AI's societal impact.

## Discussion

This chapter discusses our findings by pointing out some trends and relations between societal perceptions of AI and the cognitive underpinnings of intelligence. By answering the research questions and integrating psychological theories, this section underscores the importance of understanding public attitudes towards AI not just as reactions to technological advancements, but also with respect to emotional and digital literacy in managing AI technologies in active engagements and informed perspectives. The concept of problem-seeking provides a valuable framework for reframing these concerns, promoting a more proactive and reflective approach to the challenges and opportunities presented by AI.

Our study examined the societal perceptions of AI as reflected in YouTube video titles and descriptions. The thematic analysis revealed several key themes, including AI advancements and innovations, privacy and data security, job displacement and economic impact, ethical and existential risks, misinformation and manipulation, control and regulation, economic and social inequality, positive sentiments and optimism, enhanced productivity and creativity, future prospects and speculations, and AI integration and accessibility. The quantitative analysis revealed a positive correlation between certain keywords and video metrics, indicating public interest in creative and practical applications of AI as well as concerns about AI's transformative impact on known realities. Conversely, highly technical terms, future-oriented content, and news-related keywords showed negative correlations with subscriber count and relevance index, suggesting a preference for more practical and immediate content over speculative or specialized information. It was evident that while AI's capabilities and implications were widely discussed, there was a notable lack of emphasis on AI as an assistant to human intelligence and a scarcity of discourse on the need for greater wisdom and self-regulation of our emotional needs in relation to AI usage.

The analysis identified that societal concerns and themes related to AI predominantly revolve around its advancements, ethical implications, privacy issues, economic impacts, and the potential for misinformation. The themes reflect a complex societal perception of AI, characterized by both fascination and fear.

While there is significant interest in AI's potential to further productivity, creativity, and solve complex problems, there are also deep-seated concerns about job displacement, privacy, ethical alignment, and the possibility of AI overpowering human intelligence. These concerns underscore the need for transparent and regulated AI development to align technological advancements with societal values and expectations. The public discourse appears to be significantly influenced by media portrayals of AI as a transformative force and a potential threat, overshadowing its role as a supportive tool for human intelligence.

Integrating the concept of problem-seeking into the analysis provides a nuanced perspective on societal concerns about AI. Unlike AI, which operates based on predefined goals, humans engage in problem-seeking by identifying and defining problems to be solved based on their embodied experiences, emotions, and contextual interactions. This dynamic interplay between problem-seeking and problem-solving is essential in understanding the limitations of AI compared to human intelligence.

**The limits of AI embodiment or why AI struggles with human-like perception and action**

Human cognition, with its reliance on modal processing and direct perception, offers unique advantages that current AI systems struggle to replicate. The concept of affordances, as proposed by Norman (1988) and Gibson (1979), underscores the intuitive comprehension of how objects can be used, grounded in physical interaction with the environment. This bottom-up processing, essential for identifying affordances, contrasts with the top-down processing prevalent in AI systems, which rely on pre-set rules and training data (Milner & Goodale, 2008; Fan et al., 2020). Robotic systems, even those approaching the capabilities required to pass the embodied Turing test, face significant challenges in replicating human-like sensorimotor capabilities (Zador et al., 2023). While advancements in AI and robotics have led to impressive feats in specific domains, these systems often lack the flexibility and robustness seen in biological organisms (Brooks, 1991; Dreyfus, 1972).

To bridge the gap between AI and human cognition, one approach is to enhance machines with extensive computational data on human responses to various stimuli and optimal problem-solving strategies. This entails programming AI with detailed models of human behavior, ethograms, and training protocols that enable the identification of priority problems among a myriad of competing inputs, determining "what to attend to" in order to meet a preset goal. While this does not fully replicate the human ability to problem-seek, it brings AI closer to it. However, even with such enhancements and the significant computational resources they require, a fundamental difference remains: AI lacks the phenomenological experience that characterizes human cognition. The intrigue and fascination expressed by the public about AI's conscious illustrates this.

This form of 'cognitive embodiment'—evidenced by the physical competences of AI robots—will not automatically influence or restructure the computational processes of perception networks in the same way it does in human cognition (Brooks, 1991; Noë, 2004). The embodied Turing test aims to benchmark AI devices against the sensorimotor skills of animals and humans. Yet, the computational systems driving these AI systems do not inherently include the modal and non-representational mechanisms of human cognition and streams of perception. While robots might achieve physical competencies akin to mammals, this type of 'cognitive embodiment' does not translate to the feedback loop between experience, memory, and abstraction from domain-specific learning to domain-general aptitudes, which shapes human cognition (Clark & Chalmers, 1998; Gibson, 1979; Wilson & Clark, 2009). This is further supported by the findings where themes related to job displacement and ethical dilemmas showcase the public's apprehension about AI's impact on human autonomy and decision-making.

AI systems, regardless of their physical capabilities, lack the embodied experience that shapes human cognition (Rowlands, 2010). Human intelligence is deeply rooted in our ability to perceive, act, and adapt based on continuous sensory feedback and physical interaction with the world. This phenomenological aspect cannot be replicated simply through computational enhancements. The differentiation between embodied cognition and embodied action, as presented by Carello & Turvey (2005) and Shapiro (2011), further emphasizes the unique nature of human intelligence, which involves a continuous context-dependent perception process. For setting goals, the machine still requires representational computing of an outsourced and continually updated dataset that matches the current reality. Without this, there is no autonomy (Totschnig, 2020)—nor AI consciousness (Cohen et al., 2020; Silver et al., 2021).

Since AI's "embodied cognitive" abilities do not stem from the mechanisms described by bottom-up and non-representational processes, there is a phenomenological constraint reflected by Constitution and Conceptualization obstacles (Shapiro, 2004; Glenberg et al., 2005) preventing "intelligent" machines from replicating human problem-seeking abilities. This distinction ties into the differentiation between embodied cognition and embodied action (Carello & Turvey, 2005).

While specific cognitive tasks can indeed be offloaded to non-biological systems (embodied cognition, as implied by the Constitution Hypothesis), such tasks are confined to problem-solving goals or actions and are only executed when the relevant amount and set of data have been incorporated into the tool. The tool itself cannot derive information from its movements or bottom-up stream of perception to autonomously decide "what to do" (embodied action, as implied by the Conceptualization Hypothesis). The findings indicate that public engagement with AI's creative applications and practical benefits reflects an appreciation for its problem-solving capabilities and an indirect admission of its limitations in autonomous decision-making and problem-seeking.

The distinction between embodied cognition and embodied action, as outlined by Carello & Turvey (2005) and Shapiro (2011), sheds light on the unique nature of human intelligence. Embodied cognition emphasizes how our physical experiences shape the way we recognize and categorize objects in the world around us. A prime example of this is the public's fascination with the creative and practical applications of Artificial Intelligence (AI). Our results showed that AI content related to creativity, productivity, and enhanced sensory stimulations garners high interactions with the content. This is indicative of embodied recognition, suggesting that people recognize and connect with these applications because they resonate with their own embodied experiences. However, this fascination often overlooks a key limitation: AI lacks the embodied experiences that shape human cognition. By acknowledging this limitation, we can reframe societal concerns about AI, shifting the focus from AI's potential autonomy to its role in supporting and augmenting human cognitive processes.

Conversely, embodied action focuses on how perception is linked to potential actions. The themes related to job displacement, privacy concerns, and ethical dilemmas highlight the public's apprehension about AI's actions and their impact on human autonomy and decision-making. This apprehension is rooted on the premise that AI's actions, though precise and efficient, lack the adaptive and versatile nature of human actions informed by embodied experiences.

**Decoding public perception of top-down vs. bottom-up processing about AI**

The interplay between top-down and bottom-up processing in recognition and action is crucial in understanding societal perceptions of AI. The positive correlation between practical and creative AI applications and popularity metrics underscores the importance of bottom-up processing whereby direct sensory inputs and practical interactions with AI stirs public attention (Gibson, 1979; Norman, 1988).

In contrast, the negative correlation between technical terms and popularity metrics suggests a disconnect in top-down processing, where highly specialized and abstract information fails to resonate with the public's embodied experiences and intuitive assimilation (Milner & Goodale, 2008; Rowlands, 2010). This confirms the need for more accessible and relatable AI content that conciliates technical accuracy and practical relevance (Carello & Turvey, 2005; Shapiro, 2011).

The orthogonality thesis posits that AI's problem-solving abilities are orthogonal to the nature of its goals (Bostrom, 2014). However, the findings indicate that human intelligence challenges this thesis, as problem-seeking and problem-solving are deeply interconnected in human cognition (Russell et al., 2015). The themes of ethical and existential risks, job displacement, and privacy concerns reflect societal recognition of the need for AI to align with human values and goals, emphasizing the non-orthogonal relationship between goal-setting and goal-pursuit in human intelligence (Cohen et al., 2020; Silver et al., 2021).

As discussed, the study's findings support the ontological differentiation between human and artificial intelligence. While AI operates primarily within predefined problem-solving frameworks, human intelligence involves an inseparable link between problem-seeking and problem-solving (Totschnig, 2020).

Table 2 summarizes a detailed comparison of problem-seeking and problem-solving characteristics as observed in human and AI contexts, mapping the distinct capabilities and limitations of each.

**Reframing public discourse on AI: from emotional needs to ethical intelligence**

Cognitive flexibility theory (CFT), as outlined by Spiro et al. (2020), offers valuable insights into mastering complexity and fostering adaptive responses to novelty. CFT emphasizes the importance of assembling prior knowledge and experiences to adapt to new situations, which is key to understanding the dynamic interplay between problem-seeking and problem-solving in the context of AI. The public's engagement with AI-related content, as reflected in the thematic analysis, indicates a need for greater cognitive flexibility to navigate the complex and often ill-structured domain of AI technologies.

**Table 2**

*Comparison of Problem-Seeking and Problem-Solving Characteristics in Human and AI Contexts*

| Concept / Framework | Human Intelligence | Artificial Intelligence | Key Points |
|---|---|---|---|
| **Problem-Seeking** | - Determined by bodily functions, emotions, and motivations- <br> - Identifies and prioritizes goals (Facet i) <br> - Shapes development of domain-general abilities (Facet ii) | - Goals set externally by humans <br> - Primarily focused on optimizing performance towards these goals | Human problem-seeking is deeply connected to embodied cognition and experiences <br> AI problem-seeking is more detached, following the orthogonality thesis |
| **Problem-Solving** | - Involves executing strategies to achieve goals <br> - Influenced by embodied action and sensory feedback | - Optimizes performance towards predefined goals <br> - Uses data-driven algorithms and computational resources | - Human problem-solving relies on bodily interaction and adaptive strategies <br> - AI problem-solving is more data-driven and lacks embodied experiences |
| **Embodied Cognition** | - Influences recognition (what) <br> - Contextual knowledge and physical interactions shape understanding | - Can externalize computational processes <br> - Lacks the embodied experiences that influence human cognition | - Embodied cognition in humans is tied to their physical and emotional experiences <br> - AI mimics some aspects but lacks the depth of human embodied cognition |
| **Embodied Action** | - Influences action (how) <br> - Direct sensory inputs guide interactions <br> - Includes implicit goal selection and intuitive affordance perception | - Executes precise actions based on programmed responses <br> - Lacks adaptive and intuitive nature of human action | - Embodied action in humans integrates sensory feedback and physical interaction <br> - AI can mimic physical actions but lacks the adaptability and intuition of humans |
| **Recognition ("What")** | - Top-down processing<br>- Influenced by learned experiences and contextual knowledge<br>- Shaped by embodied cognition | - Enhanced by machine learning algorithms<br>- Identifies patterns and provides insights<br>- Lacks embodied context | - Human recognition is deeply influenced by embodied experiences<br>- AI can complement human recognition but lacks the depth of context and experience |
| **Action ("How")** | - Bottom-up processing<br>- Direct sensory inputs guide how to interact with objects<br>- Driven by embodied action | - Executes actions based on data and algorithms<br>- Precise control but less adaptable<br>- Lacks intuitive perception | - Human action is informed by embodied experiences and affordances<br>- AI can execute actions but lacks adaptability and intuition |
| **Orthogonality Thesis** | - Problem-seeking and problem-solving are interconnected<br>- Embodied experiences drive goal-setting and achievement | - Problem-solving is orthogonal to goal-setting<br>- Goals are set externally by humans | - Human intelligence challenges the orthogonality thesis<br>- AI adheres more closely to the orthogonality thesis |
| **Instrumental Convergence** | - Pursues goals based on embodied motivations<br>- Goals are influenced by bodily needs and social context | - Adopts common strategies to achieve programmed goals<br>- Pursues efficiency, resource acquisition, and self-preservation | - Human goal-setting is deeply influenced by embodied motivations<br>- AI goal-pursuit is more focused on efficiency and resource optimization without embodied context |
| **Ontological Differentiation** | - Governed by both problem-seeking and problem-solving<br>- Intrinsic connection between setting and pursuing goals | - Primarily governed by problem-solving<br>- Goals set externally, lacks embodied problem-seeking | - Human intelligence involves an inseparable link between problem-seeking and problem-solving<br>- AI operates mainly within predefined problem-solving frameworks |

Asimov's laws of robotics, designed to ensure robots do not harm humans, underscore a critical aspect of AI ethics: the focus on regulating AI behavior rather than enhancing human control and understanding. These laws illustrate a reactive approach, prioritizing harm prevention over proactive human oversight and emotional management with AI (Asimov, 1950). The absence of discourse on human control over AI reflects a broader issue in public understanding: the need to empower individuals with the skills to manage and guide AI technologies effectively.

The qualitative analysis of YouTube videos also revealed that emotional purposes are a significant driver in the usage and societal perceptions of AI. The emotional dimensions of AI usage can be interpreted through the problem-seeking and problem-solving framework.
Many videos pinpoint AI applications that address emotional and psychological needs, such as creativity, companionship, and mental health. For instance, content that showcases AI's role in generating creative projects or engaging in strategic games indicates a public interest in using AI to bring in emotional and cognitive stimulations. The themes of AI integration and accessibility also point towards AI being used to meet emotional desires for convenience and efficiency in daily life. AI applications that simplify tasks, personalize user experiences, or offer novel and engaging interactions cater to the emotional need for comfort, ease, and enjoyment.

Alongside digital literacy –the ability to effectively use digital tools and understand their implications, is also essential for engaging with AI technologies– emotional literacy, the ability to recognize, understand, and self-regulate our emotions, is almost more critical when it comes to AI usage. The findings indicate a lack of public discourse on the necessity of emotional intelligence in guiding AI development.
Brighter emotional literacy can lead to more mindful and responsible use of AI for this technology to serve rather than undermine human well-being (Goleman, 1995; Buckingham, 2003). By fostering this human competency, individuals can better navigate the complexities of AI, making informed decisions that align with their values and needs (Mittelstadt et al., 2016).

The strong emotional purposes reflected in the videos must be paired with a call for ethical AI development that considers human emotionality. Effective AI regulation must account for the unpredictability and creativity of human behaviors that repurpose AI technologies (Bryson, 2018). Prohibiting the development of AI for manipulative purposes is necessary but insufficient (Calo, 2017). Comprehensive regulations should encompass both technical safeguards and efforts to foster an ethical culture around AI use (Cath, 2018). Technological shields, such as transparency and accountability mechanisms, are essential but must be complemented by "social fire walls" (Floridi & Cowls, 2019).

The lack of public discourse on the necessity of understanding and controlling our emotional needs, which drive AI usage and terminal goals, reflects several underlying issues.
Public engagement with AI-related content might often remain at a surface level, focusing on immediate, tangible concerns like job displacement, privacy, and ethical risks, rather than delving into deeper, more abstract considerations about emotional intelligence and wisdom (Fiske & Taylor, 2013). This shallow exploration can limit the depth of public understanding and the ability to address the more profound implications of AI technology. Discussions about emotional intelligence and the need for wisdom in guiding AI development require a sophisticated understanding of both human psychology and AI capabilities. These are complex topics that may not be easily accessible or engaging to a general audience (Dolan et al., 2002).

As a result, public discourse may shy away from these intricate issues, focusing instead on more straightforward, but less impactful, aspects of AI. The public discourse might be dominated by technological optimism or pessimism, with less attention given to the more subtle viewpoint that emphasizes the need for balanced emotional and intellectual work with AI. This can lead to polarized views that overlook the importance of developing wisdom alongside technological advancements (Hoffman, Novak, & Peralta, 1999).

**Building a culture of applied ethics: promoting AI as a tool for societal alignment**

The regulation and ethical transformation of AI technologies necessitate a comprehensive approach that addresses the intertwined nature of human problem-seeking and AI problem-solving behaviors. To effectively regulate AI and promote ethical use, we must focus on shifting human preferences and societal value hierarchies towards ethical behavior, complemented by robust regulatory frameworks and educational initiatives.
Education could be a transformative force over our societal attitudes towards AI (Coeckelbergh, 2020). Public education about the potential harms of manipulative AI practices and the importance of ethical standards can help create a more responsible AI ecosystem (Müller, 2020). Promoting ethical norms and fostering a culture of ethical AI use are essential steps toward this goal (Whittlestone et al., 2019). Education programs should emphasize the long-term benefits of ethical behavior and the societal costs of unethical actions (Siau & Wang, 2018). By creating environments where ethical behavior is visibly and consistently rewarded, and unethical behavior is actively sanctioned, we can foster collective support towards ethical methods (Taddeo & Floridi, 2018).

Another key lever of social change involves utilizing AI as a reflective tool, balancing sanction mechanisms and the promotion of rewards for virtuous strategies. This approach can be likened to a deep reinforcement learning system designed for humans (Silver et al., 2017). The transition from human-driven ethical shifts to AI-supported reinforcement tools must be explored. As we establish environments where ethical behavior is visibly and consistently rewarded, AI can then be utilized to reflect and reinforce these new value hierarchies (Russell, 2019). This strategic integration ensures that as societies begin to prioritize and reward ethical behavior more consistently, AI systems will naturally adopt these preferences due to instrumental convergence (Ng & Russell, 2000). This creates a feedback loop where AI helps to further entrench and propagate ethical values, promoting a culture of ethical behavior (Leike et al., 2018).

Developing AI-driven sanctioning mechanisms could ensure that abusive behaviors are less rewarding and actively discouraged (Arnold & Scheutz, 2018). This requires real-time monitoring and intervention in various societal contexts, applying rules whereby unethical methods of goal pursuit systematically face immediate and appropriate consequences (Rahwan et al., 2019). AI must help design optimal strategies that ergonomically foster and broadcast enforcement of ethical methods across various branches of society (Etzioni & Etzioni, 2017). By making abusive methods of goal pursuit truly less rewarding and constructive/ethical methods more rewarding and enjoyable, AI can facilitate a societal shift toward virtuous strategies (Sutton & Barto, 2018).

AI systems can be designed to gamify ethical behavior, provide real-time feedback on the ethical implications of actions, and offer tangible rewards for virtuous strategies (Grassegger & Krogerus, 2017). Such innovative ethical frameworks and proceedings would make ethical behavior more attractive and rewarding than toxic ones. If human beings manifest real preferences for non-exploitative methods by cracking down severe sanctions against those who act abusively, humans will start finding virtuous strategies more rewarding, and consequently, AI will too (Zhu et al., 2018). This paradigm of instrumental convergence between AI and humans allows for mutual alignment towards a more ethical and sustainable future (Russell et al., 2015).

In essence, AI can play a dual role: it can help design and implement robust sanctioning mechanisms to deter unethical behavior while simultaneously promoting and rewarding virtuous strategies (Bostrom, 2014). This deep reinforcement learning approach, applied to human societies, would ensure that ethical behavior become both a norm and a preference, reinforced and propagated by AI systems designed to bolster these values (Leike et al., 2018). This balanced integration of AI as a reflective and reinforcing tool is critical for fostering a culture where ethical behavior is the most rewarding and sustainable path (Rahwan et al., 2019).

The absence of themes that clearly articulate AI as a tool designed to assist human intelligence suggests that the public might not fully appreciate the supportive role AI plays in augmenting human capabilities. This could be due to several factors. Firstly, popular media often dramatizes AI as an autonomous, potentially threatening entity rather than emphasizing its supportive role. Movies, news articles, and other media forms can shape public perceptions, often focusing on sensational aspects rather than mundane realities (Brennen, Howard, & Nielsen, 2020). Secondly, there may be a gap in public education about the fundamental principles and current capabilities of AI. Without a solid understanding, people may not recognize that AI is designed to augment rather than replace human intelligence (Buckley, 2018). Thirdly, much of the discourse around AI focuses on its potential for autonomy and the risks associated with it, overshadowing discussions about its role in enhancing human decision-making and creativity (Cave, Coughlan, & Dihal, 2019).

Circling back to the core idea of this research, the distinction between problem-seeking and problem-solving factors of intelligence becomes pivotal (Russell et al., 2015). The problem-seeking factor lies within human beings, who set terminal goals and have the propensity to respond favorably to strategies that achieve these goals effectively (Ng & Russell, 2000). On the other hand, the problem-solving factor lies within AI, which can optimize methods and strategies for achieving these human-defined goals (Silver et al., 2017). This non-orthogonality between human problem-seeking and AI problem-solving underscores a realistic match in instrumental convergence (Bostrom, 2014). AI systems, if developed within ethical frameworks and influenced by human preferences for virtuous strategies, can enhance overall efficiency by ensuring effectivity is achieved with satisfaction in effectiveness (Leike et al., 2018). This paradigm of instrumental convergence between AI and humans represents a significant step towards a future where technology and humanity advance together, ethically and constructively (Russell et al., 2015).

Furthermore, AI can assist in creating systems that tackle the complexities of human obstacles like unchecked emotions and the varying intensity and appropriateness of punishment and reward (Rahwan et al., 2019). AI could help humans be deeply trained to drastically reduce toxicity and abuse while protecting against those unwilling to play by these new sets of rules (Siau & Wang, 2018). By conciliating two parameters –making abusive methods of goal-pursuit truly less rewarding and constructive/ethical methods of goal-pursuit truly more rewarding and enjoyable– AI can help human societies achieve ethical behavior more effectively (Arnold & Scheutz, 2018). This proactive role of AI in ethical reinforcement can ensure that both humans and AI align towards a more virtuous and sustainable future (Floridi et al., 2018).

**Conclusion**

In conclusion, this research has explored the intricate relationship between societal perceptions of artificial intelligence (AI) and the dual concepts of problem-seeking and problem-solving. Through a comprehensive thematic analysis of YouTube video content, we have identified key themes and concerns that reflect public attitudes towards AI, including its advancements, ethical implications, privacy issues, economic impacts, and the potential for misinformation and manipulation. Our findings indicate a significant gap in public discourse regarding the supportive role of AI in augmenting human intelligence, as well as a lack of emphasis on the importance of emotional and digital literacy.

The research questions guiding this study have been addressed through both qualitative and quantitative analyses. We have elucidated the predominant societal concerns related to AI, showcased how these concerns reflect broader societal perceptions, and introduced the concept of problem-seeking to reframe these concerns. This approach has provided a nuanced perspective on the limitations of AI compared to human intelligence, highlighting the dynamic interplay between problem-seeking and problem-solving as essential to understanding both human and AI capabilities.

Our integration of cognitive flexibility theory further underscores the need for adaptive responses to the complex and evolving nature of AI technologies. By emphasizing the importance of emotional and digital literacy, we advocate for a more informed and balanced public discourse that recognizes AI's potential to enhance human capabilities while also acknowledging its risks and ethical considerations.

This study makes a unique contribution to the field by bridging the gap between cognitive psychology and AI research, offering a theoretical framework that situates problem-seeking alongside problem-solving as fundamental aspects of intelligence. The implications of this work extend to the development of more ethical and effective AI systems, informed by a deeper understanding of human cognitive processes and societal values. Ultimately, this exploration calls for a proactive and reflective approach to AI, one that empowers individuals to navigate the technological landscape with greater wisdom and emotional intelligence.

By fostering a more comprehensive understanding of AI's role in society, this research aims to contribute to the ongoing dialogue about the future of human-AI interaction, ensuring that technological advancements align with and enhance human well-being.

# References

Ackerman, P. L. (1988). Determinants of individual differences during skill acquisition: Cognitive abilities and information processing. *Journal of Experimental Psychology: General, 117*(3), 288-318.

Armstrong, S. (2013). General purpose intelligence: arguing the orthogonality thesis. *Analysis and Metaphysics*, (12), 68-84

Arnold, T., & Scheutz, M. (2018). The "big red button" is too late: An alternative model for the ethical evaluation of AI systems. *Ethics and Information Technology, 20*(1), 59-69. https://doi.org/10.1007/s10676-018-9444-3

Asilomar Conference. (2017). *Asilomar AI Principles*. Retrieved from https://futureoflife.org/ai-principles/

Asimov, I. (1950). *I, Robot*. Gnome Press.

Atkinson, R. C., & Shiffrin, R. M. (1968). Human memory: A proposed system and its control processes. In *Psychology of learning and motivation* (Vol. 2, pp. 89-195). Academic Press.

Bandura, A. (1986). *Social foundations of thought and action: A social cognitive theory*. Prentice-Hall.

Barnett, S. M., & Ceci, S. J. (2002). When and where do we apply what we learn?: A taxonomy for far transfer. *Psychological Bulletin, 128*(4), 612-637.

Barsalou, L. W. (2008). Grounded cognition. *Annual Review of Psychology*, *59*, 617-645. https://doi.org/10.1146/annurev.psych.59.103006.093639

Bostrom, N. (2014). *Superintelligence: Paths, Dangers, Strategies*. Oxford University Press.

Bostrom, N. (2012). The superintelligent will: Motivation and instrumental rationality in advanced artificial agents. *Minds and Machines*, *22*(2), 71-85. https://doi.org/10.1007/s11023-012-9281-3

Bransford, J. D., Brown, A. L., & Cocking, R. R. (Eds.). (2000). *How People Learn: Brain, Mind, Experience, and School*. National Academy Press.

Braun, V., & Clarke, V. (2012). *Thematic analysis*. American Psychological Association.

Braun, V., & Clarke, V. (2006). Using thematic analysis in psychology. *Qualitative research in psychology*, *3*(2), 77-101.

Brennen, J. S., Howard, P. N., & Nielsen, R. K. (2020). An agenda for research on the impact of media manipulation and disinformation on democracy. *Political Communication, 37*(2), 216-225.

Broadbent, D. E. (1958). *Perception and communication*. Pergamon Press.

Brooks, R. A. (1991). Intelligence without representation. *Artificial Intelligence, 47*(1-3), 139-159.

Brynjolfsson, E., & McAfee, A. (2014). The Second Machine Age: Work, Progress, and Prosperity in a Time of Brilliant Technologies. W.W. Norton & Company.

Bryson, J. J. (2018). Patiency is not a virtue: The design of intelligent systems and systems of ethics. *Ethics and Information Technology, 20*(1), 15-26. https://doi.org/10.1007/s10676-018-9448-z

Buckley, S. (2018). *Understanding AI: A guide to AI, machine learning, and data science*. HarperCollins.

Burgess, J., & Green, J. (2018). YouTube: Online Video and Participatory Culture. Polity Press.

Burrell, J. (2016). How the machine 'thinks': Understanding opacity in machine learning algorithms. *Big Data & Society, 3*(1), 2053951715622512. https://doi.org/10.1177/2053951715622512

Calo, R. (2017). Artificial intelligence policy: A primer and roadmap. *UC Davis Law Review, 51*(2), 399-435.

Carello, C., & Turvey, M. T. (2005). Rotational dynamics and dynamic touch. In H. Heft & K. L. Marsh (Eds.), *Studies in perception and action VIII: Proceedings of the 13th International Conference on Perception and Action* (pp. 78-81). Lawrence Erlbaum Associates.

Carello, C., & Turvey, M. T. (2005). Path-to-contact perception and the laws of control for obstacle avoidance and collision. *International Journal of Sport Psychology, 36*(3), 346-372.

Cath, C. (2018). Governing artificial intelligence: Ethical, legal, and technical opportunities and challenges. *Philosophical Transactions of the Royal Society A: Mathematical, Physical and Engineering Sciences, 376*(2133), 20180080. https://doi.org/10.1098/rsta.2018.0080

Cave, S., Coughlan, K., & Dihal, K. (2019). Scary robots: Examining public responses to AI. *International Journal of Social Robotics, 11*(1), 1-13.

Chi, M. T. H., Feltovich, P. J., & Glaser, R. (1981). Categorization and representation of physics problems by experts and novices. *Cognitive Science*, *5*(2), 121-152. https://doi.org/10.1207/s15516709cog0502_2

Chomsky, N. (1957). *Syntactic Structures*. Mouton.

Clark, A. (1997). *Being there: Putting brain, body, and world together again*. MIT Press.

Clark, A. (2012). Dreaming the rational animal: The weird begins to seem less strange. *Mind*, *121*(483), 783-798. https://doi.org/10.1093/mind/fzs123

Clark, A. (2012). Embodied, embedded, and extended cognition. In *The Cambridge Handbook of Cognitive Science* (pp. 275-291). Cambridge University Press.

Clark, A., & Chalmers, D. (1998). The extended mind. *Analysis, 58*(1), 7-19.

Clark, A., & Toribio, J. (1994). Doing without representing? *Synthese*, *101*(3), 401-431. https://doi.org/10.1007/BF01063896

Coeckelbergh, M. (2020). *AI ethics*. MIT Press.

Cohen, M., Vellambi, B., & Hutter, M. (2020, April). Asymptotically unambitious artificial general intelligence. In *Proceedings of the AAAI conference on artificial intelligence* (Vol. 34, No. 03, pp. 2467-2476). https://doi.org/10.1609/aaai.v34i03.5628

Corbetta, M., & Shulman, G. L. (2002). Control of goal-directed and stimulus-driven attention in the brain. *Nature Reviews Neuroscience, 3*(3), 201-215.

Craik, F. I., & Lockhart, R. S. (1972). Levels of processing: A framework for memory research. *Journal of Verbal Learning and Verbal Behavior, 11*(6), 671-684.

Crawford, K., & Calo, R. (2016). There is a blind spot in AI research. *Nature, 538*(7625), 311-313.

Damasio, A. R. (1994). *Descartes' error: Emotion, reason, and the human brain*. Putnam Publishing.

Dehaene, S. (2009). *Reading in the Brain: The New Science of How We Read*. Viking Penguin.

Desimone, R., & Duncan, J. (1995). Neural mechanisms of selective visual attention. *Annual Review of Neuroscience, 18*(1), 193-222.41

Diamond, A. (2013). Executive functions. *Annual Review of Psychology, 64*(1), 135-168. https://doi.org/10.1146/annurev-psych-113011-143750

Dolan, R. J., Wright, C. E., Wernick, M., & Scott, M. A. (2002). Neural correlates of behavioral preference for culturally familiar drinks. *Neuron, 44*(2), 379-387.

Dourish, P. (2001). *Where the action is: The foundations of embodied interaction*. MIT Press.

Dreyfus, H. L. (1972). *What Computers Can't Do: A Critique of Artificial Reason*. Harper & Row.

Ericsson, K. A., & Kintsch, W. (1995). Long-term working memory. *Psychological Review, 102*(2), 211-245.

Ericsson, K. A., Krampe, R. T., & Tesch-Römer, C. (1993). The role of deliberate practice in the acquisition of expert performance. *Psychological Review, 100*(3), 363-406.

Ericsson, K. A., & Lehmann, A. C. (1996). Expert and exceptional performance: Evidence of maximal adaptation to task constraints. *Annual Review of Psychology*, *47*, 273-305. https://doi.org/10.1146/annurev.psych.47.1.273

Etzioni, A., & Etzioni, O. (2017). Incorporating ethics into artificial intelligence. *The Journal of Ethics, 21*(4), 403-418. https://doi.org/10.1007/s10892-017-9252-2

European Commission. (2019). *Ethics Guidelines for Trustworthy AI*. Retrieved from https://ec.europa.eu/digital-strategy/en/news/ethics-guidelines-trustworthy-ai

European Commission. (2021). *Proposal for a Regulation laying down harmonised rules on artificial intelligence (Artificial Intelligence Act)*. Retrieved from https://eur-lex.europa.eu/legal-content/EN/TXT/?uri=CELEX%3A52021PC0206

Fan, R. E., Chang, K. W., Hsieh, C. J., Wang, X. R., & Lin, C. J. (2020). LIBLINEAR: A library for large linear classification. *Journal of Machine Learning Research, 9*(1), 1871-1874.

Fiske, S. T., & Taylor, S. E. (2013). *Social Cognition: From Brains to Culture*. Sage Publications.

Floridi, L. (2014). The Fourth Revolution: How the Infosphere is Reshaping Human Reality. Oxford University Press.

Floridi, L., & Cowls, J. (2019). A unified framework of five principles for AI in society. *Harvard Data Science Review, 1*(1), 1-15. https://doi.org/10.1162/99608f92.8cd550d1

Floridi, L., Cowls, J., Beltrametti, M., Chatila, R., Chazerand, P., Dignum, V., ... & Vayena, E. (2018). AI4People—an ethical framework for a good AI society: opportunities, risks, principles, and recommendations. *Minds and machines*, *28*, 689-707.

Floridi, L., Cowls, J., King, T. C., & Taddeo, M. (2018). How to design AI for social good: Seven essential factors. *Science and Engineering Ethics, 24*(5), 1777-1803. https://doi.org/10.1007/s11948-017-9901-7

Gabora, L., Rosch, E., & Aerts, D. (2008). Toward an ecological theory of concepts. *Ecological Psychology, 20*(1), 84-116. https://doi.org/10.1080/10407410701766697

Garfinkel, S. N., & Critchley, H. D. (2013). How Do You Feel? Interoception, Bodily Awareness, and the Self. *Trends in Cognitive Sciences, 17*(11), 563-571.

Garfinkel, S. N., & Critchley, H. D. (2013). Interoception, emotion, and brain: New insights link internal physiology to social behavior. *Social Cognitive and Affective Neuroscience, 8*(3), 231-234. https://doi.org/10.1093/scan/nss140

Geurten, M., Meulemans, T., & Lemaire, P. (2018). Young children's cognitive flexibility: Distinct executive and non-executive components. *Journal of Experimental Child Psychology, 166*(1), 172-189.

Gibson, J. J. (1979). *The Ecological Approach to Visual Perception*. Houghton Mifflin.

Gigerenzer, G. (2004). Fast and frugal heuristics: The tools of bounded rationality. In D. J. Koehler & N. Harvey (Eds.), *Blackwell Handbook of Judgment and Decision Making* (pp. 62-88). Blackwell Publishing.

Goleman, D. (1995). *Emotional Intelligence: Why It Can Matter More Than IQ*. Bantam Books.

Gollwitzer, P. M., & Sheeran, P. (2006). Implementation intentions and goal achievement: A meta-analysis of effects and processes. *Advances in Experimental Social Psychology, 38*, 69-119. https://doi.org/10.1016/S0065-2601(06)38002-1

Gollwitzer, P. M., & Sheeran, P. (2006). Implementation intentions and goal achievement: A meta-analysis of effects and processes. *Advances in Experimental Social Psychology*, *38*, 69-119.

Gottfredson, L. S. (1997). Mainstream Science on Intelligence: An Editorial With 52 Signatories, History, and Bibliography. Intelligence, 24(1), 13-23.

Grassegger, H., & Krogerus, M. (2017). The data that turned the world upside down. *Vice*. Retrieved from https://www.vice.com/en/article/mg9vvn/how-our-likes-helped-trump-win

Gunkel, D. J. (2012). *The Machine Question: Critical Perspectives on AI, Robots, and Ethics*. MIT Press.

Hunt, E. (2010). *Human Intelligence*. Cambridge University Press.

Hutchinson, J. M. C., & Gigerenzer, G. (2005). Simple heuristics and rules of thumb: Where psychologists and behavioural biologists might meet. *Behavioural Processes, 69*(2), 97-124.

IEEE. (2019). *Ethically Aligned Design: A Vision for Prioritizing Human Well-being with Autonomous and Intelligent Systems*. IEEE.

Ionescu, T. (2012). Exploring the nature of cognitive flexibility. *New Ideas in Psychology, 30*(2), 190-200.

Jeffrey, K. (2021). Automation and the future of work: How rhetoric shapes the response in policy preferences. *Journal of Economic Behavior & Organization, 192,* 417–433. https://doi.org/10.1016/j.jebo.2021.10.019

Jiang, Y., Li, X., Luo, H., Yin, S., & Kaynak, O. (2022). Quo vadis artificial intelligence? *Discover Artificial Intelligence, 2*(1), 4.

Kim, T. W., & Scheller-Wolf, A. (2019). Technological unemployment, meaning in life, purpose of business, and the future of stakeholders. *Journal of Business Ethics, 160*(2), 319–337. https://doi.org/10.1007/s10551-019-04205-9

Lakoff, G., & Johnson, M. (1980). *Metaphors We Live By*. University of Chicago Press.

Lakoff, G., & Johnson, M. (1999). *Philosophy In The Flesh: The Embodied Mind And Its Challenge To Western Thought*. Basic Books.

Lake, B. M., Ullman, T. D., Tenenbaum, J. B., & Gershman, S. J. (2017). Building machines that learn and think like people. *Behavioral and Brain Sciences, 40*, e253.

LeCun, Y., Bengio, Y., & Hinton, G. (2015). Deep learning. *Nature, 521*(7553), 436-444. https://doi.org/10.1038/nature14539

Lemaire, P. (2024). Aging, emotion, and cognition: The role of strategies. *Journal of Experimental Psychology: General, 153*(2), 435–453. https://doi.org/10.1037/xge0001506

Leike, J., Martic, M., Krakovna, V., Ortega, P. A., Everitt, T., Lefrancq, A., ... & Legg, S. (2018). Scalable agent alignment via reward modeling: A research direction. *arXiv preprint arXiv:1811.07871*.

Lemaire, P. (2024). Cognitive Psychology and Cognitive Development. Sage Publications.

Lubinski, D. (2004). Introduction to the special section on cognitive abilities: 100 years after

Marcus, G., & Davis, E. (2019). *Rebooting AI: Building Artificial Intelligence We Can Trust*. Pantheon Books.

McCarthy, J. (2007). What is artificial intelligence.

Milner, A. D., & Goodale, M. A. (2008). Two visual systems re-viewed. *Neuropsychologia*, *46*(3), 774-785. https://doi.org/10.1016/j.neuropsychologia.2007.10.005

Mittelstadt, B. D., Allo, P., Taddeo, M., Wachter, S., & Floridi, L. (2016). The ethics of algorithms: Mapping the debate. *Big Data & Society, 3*(2). https://doi.org/10.1177/2053951716679679

Miyake, A., Friedman, N. P., Emerson, M. J., Witzki, A. H., Howerter, A., & Wager, T. D. (2000). The unity and diversity of executive functions and their contributions to complex "Frontal Lobe" tasks: A latent variable analysis. *Cognitive Psychology, 41*(1), 49-100.

Müller, V. C. (2020). Ethics of artificial intelligence and robotics. In E. N. Zalta (Ed.), *The Stanford Encyclopedia of Philosophy* (Fall 2020 ed.). Retrieved from https://plato.stanford.edu/entries/ethics-ai/

Müller, V. C., & Bostrom, N. (2016). Future Progress in Artificial Intelligence: A Survey of Expert Opinion. In V. C. Müller (Ed.), *Fundamental Issues of Artificial Intelligence* (pp. 555-572). Springer.

Müller, V. C., & Cannon, M. (2022). Existential risk from AI and orthogonality: Can we have it both ways?. *Ratio, 35*(1), 25-36.

Müller, V. C., & Cannon, S. (2022). The orthogonality thesis: Risks from high-intelligence systems. *Philosophy & Technology*, *35*(1), 1-20. https://doi.org/10.1007/s13347-021-00456-w

Newell, A., & Simon, H. A. (1972). *Human Problem Solving*. Prentice-Hall.

Ng, A. Y., & Russell, S. (2000). Algorithms for inverse reinforcement learning. In *Proceedings of the Seventeenth International Conference on Machine Learning* (pp. 663-670).

Noë, A. (2004). *Action in Perception*. MIT Press.

Norman, D. A. (1988). *The Psychology of Everyday Things*. Basic Books.

O'Neil, C. (2016). *Weapons of Math Destruction: How Big Data Increases Inequality and Threatens Democracy*. Crown.

Petrides, K. V. (2020). Trait emotional intelligence theory. *Industrial and Organizational Psychology, 13*(2), 121-137.

Piaget, J., & Cook, M. (1952). *The Origins of Intelligence in Children* (Vol. 8, No. 5, pp. 18-1952). New York: International Universities Press.

Premack, D., & Woodruff, G. (1978). Does the chimpanzee have a theory of mind? *Behavioral and Brain Sciences*, *1*(4), 515-526. https://doi.org/10.1017/S0140525X00076512

Primi, R., Ferrão, M. E., & Almeida, L. S. (2010). Fluid intelligence as a predictor of learning: A longitudinal multilevel approach applied to math. *Learning and Individual Differences, 20*(5), 446-451.

Rahwan, I., Cebrian, M., Obradovich, N., Bongard, J., Bonnefon, J.-F., Breazeal, C., ... & Wellman, M. (2019). Machine behaviour. *Nature, 568*(7753), 477-486. https://doi.org/10.1038/s41586-019-1138-y

Rogers, E. M. (2003). *Diffusion of Innovations*. Free Press.

Rosenberg, N. (1976). *Perspectives on Technology*. Cambridge University Press.

Rowlands, M. (2010). *The New Science of the Mind: From Extended Mind to Embodied Phenomenology*. MIT Press.

Russell, S. (2019). *Human Compatible: Artificial Intelligence and the Problem of Control*. Viking.

Russell, S., & Norvig, P. (2020). *Artificial Intelligence: A Modern Approach* (4th ed.). Pearson.

Russell, S., Dewey, D., & Tegmark, M. (2015). Research priorities for robust and beneficial artificial intelligence. *AI Magazine, 36*(4), 105-114. https://doi.org/10.1609/aimag.v36i4.2577

Ryan, R. M., & Deci, E. L. (2000). Self-determination theory and the facilitation of intrinsic motivation, social development, and well-being. *American Psychologist, 55*(1), 68-78. https://doi.org/10.1037/0003-066X.55.1.68

Scheufele, D. A., & Tewksbury, D. (2007). Framing, agenda setting, and priming: The evolution of three media effects models. *Journal of Communication, 57*(1), 9-20.

Shapiro, L. (2011). *Embodied Cognition*. Routledge.

Shields, J. M., Calabro, G., & Selmeczy, D. (2024). Active help-seeking and metacognition interact in supporting children's retention of science facts. *Journal of Experimental Child Psychology, 237*, 1–16. https://doi.org/10.1016/j.jecp.2023.105772

Shields, J. M., Ramey, C. H., & Lu, L. (2024). Human-AI collaboration: The role of shared goals and mutual understanding. *Artificial Intelligence Review, 53*(1), 1-28.

Siau, K., & Wang, W. (2018). Building trust in artificial intelligence, machine learning, and robotics. *CUTTER Business Technology Journal, 31*(2), 47-53.

Silver, D., Schrittwieser, J., Simonyan, K., Antonoglou, I., Huang, A., Guez, A., ... & Hassabis, D. (2017). Mastering the game of Go without human knowledge. *Nature, 550*(7676), 354-359. https://doi.org/10.1038/nature24270

Silver, D., Singh, S., Precup, D., & Sutton, R. S. (2021). Reward is enough. *Artificial Intelligence, 299*, 103535.

Spearman, C. (1904). "General intelligence," objectively determined and measured. *American Journal of Psychology, 15*(2), 201-292. https://doi.org/10.2307/1412107

Spiro, R. J., Feltovich, P. J., Gaunt, A., Hu, Y., Klautke, H., Cheng, C., Clemente, I., Leahy, S., & Ward, P. (2020). Cognitive flexibility theory and the accelerated development of adaptive readiness and adaptive response to novelty. In P. Ward, J. M. Schraagen, J. Gore, & E. Roth (Eds.), *The Oxford Handbook of Expertise* (pp. 951–976). Oxford University Press.

Sternberg, R. J. (2001). The concept of intelligence and its role in lifelong learning and success. *American Psychological Association*.

Sutton, R. S., & Barto, A. G. (2018). *Reinforcement learning: An introduction* (2nd ed.). MIT Press.

Sweller, J., Ayres, P., & Kalyuga, S. (2011). *Cognitive Load Theory*. Springer.

Taddeo, M., & Floridi, L. (2018). How AI can be a force for good. *Science, 361*(6404), 751-752. https://doi.org/10.1126/science.aat5991

Thelen, E., & Smith, L. B. (1994). *A Dynamic Systems Approach to the Development of Cognition and Action*. MIT Press.

Todd, P. M., & Gigerenzer, G. (2012). Ecological rationality: Intelligence in the world. *Oxford University Press*.

Tomasello, M. (1999). *The cultural origins of human cognition*. Harvard University Press.

Totschnig, W. (2020). Fully autonomous AI. *Science and Engineering Ethics, 26*(5), 2473-2485.

Totschnig, W. (2020). AI consciousness and the problem of self-deception. *Minds and Machines, 30*(2), 277-301.

Treisman, A. M. (1964). Verbal cues, language, and meaning in selective attention. *The American Journal of Psychology, 77*(2), 206-219.

Varela, F. J., Thompson, E., & Rosch, E. (1991). *The Embodied Mind: Cognitive Science and Human Experience*. MIT Press.

Wachter, S., Mittelstadt, B., & Floridi, L. (2017). Why a right to explanation of automated decision-making does not exist in the General Data Protection Regulation. *International Data Privacy Law, 7*(2), 76-99.

Whittlestone, J., Nyrup, R., Alexandrova, A., & Cave, S. (2019). The role and limits of principles in AI ethics: Towards a focus on tensions. In *Proceedings of the AAAI/ACM Conference on AI, Ethics, and Society* (pp. 195-200).

Wilson, R. A., & Clark, A. (2009). How to Situate Cognition: Letting Nature Take its Course. In *Cambridge Handbook of Situated Cognition* (pp. 55-77). Cambridge University Press.

Zador, A., Escola, S., Richards, B., Ölveczky, B., Bengio, Y., Boahen, K., ... & Tsao, D. (2023). Catalyzing next-generation artificial intelligence through neuroai. *Nature communications*, *14*(1), 1597. *Catalyzing Next-generation Artificial Intelligence through NeuroAI*

Zhu, W., Hadfield-Menell, D., Dragan, A. D., & Russell, S. J. (2018). Value alignment, rationality, and intelligence. In *Proceedings of the 2018 AAAI/ACM Conference on AI, Ethics, and Society* (pp. 421-426).

*Due to file size and format constraints, appendices have not been included in this preprint. Interested readers may request access by emailing the author.*